\documentclass[a4paper,11pt]{paper}
\usepackage{cite}
\usepackage{bm}
\usepackage{amsfonts,amsmath}
\usepackage{amssymb}
\usepackage[normalem]{ulem}
\usepackage[dvips]{graphicx}
\usepackage{scalefnt}
\begin{document}

\author{Lopes,~L.~L.~and~Menezes,~D.~P.}
\title{Neutron Stars with Hyperons subject to Strong Magnetic Field}

\maketitle

\abstract{Neutron stars are one of the most exotic objects in the universe and a unique laboratory to study
 the nuclear matter above the nuclear saturation density. In this work, we study the equation of state of the
 nuclear matter within a relativistic model subjected to a strong magnetic field. We then apply this EoS to study
 and describe some of the physical characteristics of neutron star, especially the mass-radius relation
 and chemical compositions.  To study the influence of a the magnetic field and the hyperons in the stellar
 interior, we consider altogether four solutions: two different values of magnetic field to obtain a weak
 and a strong influence, and two configurations: a family of neutron stars formed only by protons,
 electrons and neutrons and a family formed by protons, electrons, neutrons, muons and hyperons. The limit and the
 validity of the results found are discussed with some care.  In all cases  the particles that constitute
 the neutron star are in $\beta$ equilibrium and zero total net charge. Our work indicates that the effect of
 a strong magnetic field has to  be taken into account in the description of magnetars, mainly if we believe 
that there are hyperons in their interior, in which case, the influence of the magnetic field can increase the
 mass by more than $10\%$. We have also seen that although a  magnetar can reach 
2.48$M_{\odot}$, a natural explanation of why we do not  know pulsars with masses above 2.0$M_{\odot}$ arises.
We also discuss how the magnetic field affects the strangeness fraction in some standard neutron star masses and, to conclude
our paper, we revisit the direct URCA process related to the cooling of the
neutron stars and show how it is affected by the hyperons and
 the magnetic field.}
 
 \noindent
{\bf Keywords:}

\noindent Neutron stars, pulsars, strong magnetic field, hyperons.

\section{Introduction}
\label{intro}
Neutron stars are compact objects maintained by the equilibrium of gravity and the degenerescence pressure of the fermions
 together with a strong nuclear repulsion force due to the high density reached  in their interior.  Since we do not know
 yet the precise and detailed structure and composition of the inner core of a neutron star, many models have been used to
 describe it. In the literature we can find some standard ones:  hadronic neutron stars,  quark stars, strange stars and
 hybrid stars~\cite{Glen,shap,Max}.

In the present work we study a hadronic neutron star constituted by nucleons and hyperons and subject to a strong magnetic field.
 The presence of hyperons is justifiable since the constituents of neutron stars are fermions. So, according to the Pauli Principle,
 as the baryon density increases, so do the Fermi momentum and the Fermi energy. Ultimately the Fermi energy  exceeds the masses of
 the heavier baryons~\cite{Glen}. On the other hand, some strange objects like the soft gamma-ray repeaters and anomalous X-ray
 pulsars can be explained assuming that these objects are neutron stars subject to a strong magnetic fields on their  surface.
 These objects are called magnetars~\cite{Duncan}. Although the magnetic field of the  magnetars do not exceed $10^{15}G$ in
 their surface, it is well-accepted in the literature that the magnetic field in the core of the neutron stars can reach values
 greater than $10^{18}G$~\cite{Pal,Pal2}.  Due to the large densities in the neutron star interior, we do not expect any significant
 influence of the magnetic field till it reaches values of the order of $10^{18}G$. Besides, it is  well-established in the literature
 that the direct URCA process is an efficient method to cool neutron stars if the proton fraction reaches values of $11 \sim 15\%$~\cite{LPPH}.
 Hence, the study of the proton fraction in the interior of the neutron stars is very important to determine how fast is the
 cooling.

This paper is organized as follows: we make a  review of the formalism of the non-linear Walecka model (NLWM)  in the presence of
 a magnetic field. Then we present the numerical results showing how the presence of hyperons and a strong magnetic field affects
 the EoS and the chemical composition. We study how these terms alter the macroscopic mass-radius relation of the neutron stars
 and compare our results with those found in literature. To conclude our paper we discuss how the magnetic field affects
 the strangeness fraction for some standard masses, and revisit the direct URCA process.

\section{The Formalism}
\label{sec:1}

The total Lagrangian is given by~\cite{Ra2,Bro}:

\begin{equation}
\mathcal{L} =  \sum_b \mathcal{L}_b  + \mathcal{L}_m  + \sum_l\mathcal{L}_l + \mathcal{L_B},  \label{1}
\end{equation}
where $b$ stands for the baryons,  $m$ for the mesons,  $l$  for the leptons, and $\mathcal{B}$ for the electromagnetic field itself.
 The sum in $b$ can run over the eight lighter baryons and in $l$ over the two lighter leptons. Explicitly, in the presence of a
 electromagnetic field  the Lagrangian is:

\begin{equation} 
\mathcal{L}_b = \bar{\Psi_b }[\gamma_u(i\partial^{\mu} - eA^{\mu} - g_{v,b}\omega^{\mu} - g_{\rho, b} I_{3b}  \rho^{\mu} ) - (M_b - g_{s,b} \sigma) ] \Psi_b , \label{2}
\end{equation}

\begin{eqnarray}
\mathcal{L}_m  = \frac{1}{2} \partial_\mu \sigma \partial^{\mu} \sigma - \frac{1}{2} m_s^2 \sigma^2 + \frac{1}{2} m_v^2\omega_\mu \omega^\mu - \frac{1}{4}\Omega_{\mu \nu}\Omega^{\mu \nu} +  \nonumber \\
+  \frac{1}{2} m_\rho^2 (\rho_\mu \rho^{ \mu}) - \frac{1}{4} {P}_{\mu \nu}  {P}^{\mu \nu}   -\frac{1}{3!}\kappa\sigma ^3 - \frac{1}{4!}\lambda\sigma ^4    \label{4} , \nonumber \\
\end{eqnarray}

\begin{equation}
\mathcal{L}_l =  \bar{\psi_l}[\gamma_u(i\partial^\mu - eA^{\mu})] - m_l]\psi_l , \label{3}
\end{equation}

\begin{equation}
\mathcal{L_B} = - \frac{1}{16 \pi}F_{\mu \nu} F^{\mu \nu} , \label{4e1}
\end{equation}
where $\Psi_b$ and $\psi_l$ are the baryon and lepton Dirac fields, respectively. The baryon mass and  isospin projection  are denoted by
 $M_b$ and ${I_{3b}}$ respectively. The masses of the leptons are $m_l$ and the electric charge of the particles is given by $e$. The  antisymmetric
 mesonic and electromagnetic  field strength tensors are given by their usual expressions:
  $\Omega_{\mu \nu}$ = $\partial_u \omega_\nu - \partial_\nu \omega_\mu$, $P_{\mu \nu}$ = $\partial_\mu \rho_\nu - \partial_\nu \rho_\mu $ and
  $F_{\mu \nu}$ = $\partial_\mu A_\nu - \partial_\nu A_\mu$. The $\gamma_\mu$ are the Dirac matrices~\cite{griff}. The strong interaction couplings
 are denoted by $g$,  and the meson masses by $m$, all with appropriate subscripts. The second subscript at the $g$ constant is due to the distinctive
 coupling of hyperons with the mesons. In this work we assume that  $g_{s,H}$ = 0.7$g_{s, N}$;  $g_{v,H}$ = 0.783$g_{v, N}$ and
 $g_{\rho, H}$ = 0.783$g_{\rho, N}$~\cite{deb2}, where $H$ denotes hyperons and $N$ nucleons. The hadronic part of the Lagrangian is the called NLWM.
 The leptons are  included in the total Lagrangian density as a non-interacting Fermi gas in order to account the $\beta$ equilibrium in the star.

To solve the equation of motion, we use the mean field approximation, where the meson fields are replaced by their expectation values, \textit{i.e.}:
  $\sigma$ $\to$ $\left < \sigma \right >$ = $\sigma_0$,   $\omega^\mu$ $\to$ $\delta_{0 \mu}\left <\omega^\mu  \right >$ = $\omega_{0}$  and
   $\rho^\mu$ $\to$ $\delta_{0 \mu}\left <\rho^\mu  \right >$ = $\rho_{0}$.

In this work we use a GM1 parametrization~\cite{Glen3}, which can describe the most important properties of nuclear matter and reproduce the macroscopic
 properties of the neutron stars consistent with those observed in nature. The GM1 parameters are showed in table \ref{tab:1}.

\begin{table*}[ht]
\centering

\caption{Values of GM1 parametrization.}
\label{tab:1}       
\begin{tabular}{llllll}\hline\noalign{\smallskip}
Set & $(g_s/m_s)^2$ & $(g_v/m_v)^2$ & $(g_\rho/m_\rho)^2$ & $\kappa/M_N$ & $\lambda$  \\
\noalign{\smallskip}\hline\noalign{\smallskip}
GM1 & 11.785 $fm^2$& 7.148 $fm^2$ & 4.410 $fm^2$ & 0.005894 & -0.006426 \\
\noalign{\smallskip}\hline
\end{tabular}
\vspace*{1cm}  
\end{table*}

This parametrization is fixed so that the incompressibility of nuclear matter $K$ = 300 MeV, and  the nuclear saturation density $n_0$ = 0.153$fm^{-3}$.
 The masses of the baryon octet are $M_N$ = 939 MeV (nucleons), $M_{\Lambda}$ = 1116 MeV, $M_{\Sigma}$ =1193 MeV and $M_{\Xi}$ = 1318 MeV.  The meson
 masses are $m_s$ = 400 MeV, $m_v$ = 783 MeV and $m_\rho$ = 770 MeV. The masses of the leptons are $m_e$ = 0.511 MeV and $m_\mu$ = 105.66 MeV.
 Applying the Euler-Lagrange in equation (\ref{1}) in the absence of an electric field, the equation of motion in the mean field approximation for an
 arbitrary baryon becomes:

 \begin{equation}
 [ \gamma_0   (i \partial^0  - g_{v, b}\omega_{0} - g_{\rho, b} I_{3b}  \rho_0   ) - \gamma_j \; (i \partial^j - eA^j) - M^{*}_b   ] \Psi \;  =0,   \label{5}
\end{equation}
where

\begin{equation}
M^{*}_b = M_b - g_{s, b}\sigma_0 , \label{6}
\end{equation}
is the baryon effective mass.

\noindent For an uncharged particle  $eA^\mu$ is always zero. The  quantization rules are:   $i\partial^{0}$ = $E$ and   $i\partial^{j}=k^{j}$, where $k^{j}$ is the
 momentum in $j$ direction. Setting  $E - g_{v, b}\omega_0 - g_{\rho, b} I_{3b} \rho_0 = E^{*}$, we have the following equation of motion written in a block matrix:

\begin{equation}
\left(\begin{array}{rr}

            (E^{*} -  M^{*}_b)  & - \;  \mathbf{\sigma}.\mathbf{k} \\
          \;  \mathbf{\sigma}.\mathbf {k} & - \; (E^{*} \rho_0 + M^{*}_b) 
            
\end{array}\right)
\left(\begin{array}{r}
u_{A}\\
u_{B}

\end{array}\right)=0. \label{7}
\end{equation}

This is an eigenvalue equation, which can be solved as the free  Dirac equation for an effective mass and energy, whose  solution is:

\begin{equation}
E = \sqrt{k^2 + M^{*2}_b} + g_{v, b}\omega_0 + g_{\rho, b} I_{3b} \rho_0 =  \mu, \label{8}
\end{equation}
where $\mu$ is the chemical potential.

For a charged  baryon, the Dirac equation  assumes the following form:

 \begin{equation}
\left(\begin{array}{rr}

            (E^{*} - M^{*}_b)  & - \;  \mathbf{\sigma}.(\mathbf{k} -e\mathbf{A}) \\
          \;  \mathbf{\sigma}.(\mathbf {k} -e\mathbf{A}) & - \; (E^{*} - M^{*}_b) 
            
\end{array}\right)
\left(\begin{array}{r}
u_{A}\\
u_{B}

\end{array}\right)=0. \label{9}
\end{equation}
To produce a constant magnetic field in the $z$ direction we choose:  $A_2$ = $A_3$ = 0, $A_1 = - B$y. The solution of this eigenvalue equation
 is well-know in the literature~\cite{Pal,Ra2,Ra}:

\begin{equation}
E^{*} = \sqrt{ M^{*2}_b +  k_z^2 +2\nu |e|B} , \nonumber
\end{equation}
\begin{equation}
E = \sqrt{ M^{*2}_b +  k_z^2 +2\nu |e|B} +g_{v, b}\omega_0 + g_{\rho, b}I_{3b} \rho_0 = \mu , \label{10}
\end{equation}
where the discrete parameter  $\nu$ is called Landau level ($LL$)  and  $\mu$ is the chemical potential.  The first $LL$, $\nu=0$, is non degenerate
 and  all the others are two-fold degenerate.  For the leptons, since they don't feel the strong force:

\begin{equation}
E_l = \sqrt{ m^{2}_l +  k_z^2 +2\nu |e|B} = \mu_l . \label{11}
\end{equation}

The expected values for the mesons are:

\begin{equation}
\omega_0  =  \sum_{ub} \frac{g_{v,b}}{m_v^2}n^{ub} +  \; \sum_{cb}\frac{g_{v, b}}{m_v^2} n^{bc},  \label{12}
\end{equation}

\begin{equation}
 \sigma_0 = \sum_{ub} \frac{g_{s,b}}{m_s^2} n_s^{ub} + \; \sum_{cb} \frac{g_{s,b}}{m_s^2} n_s^{bc} - \;  \frac{1}{2}\frac{\kappa}{m_s^2} \sigma_0^2 - \frac{1}{6}\frac{ \lambda}{m_s^2} \sigma_0^3 , \label{13}
 \end{equation}

\begin{equation}
\rho_0 =   \sum_{ub} \frac{g_{\rho, b}}{m_\rho^2} n^{ub}I_{3b} + \sum_{cb} \frac{g_{\rho,b}}{m_\rho^2} n^{cb} I_{3b} , \label{14}
\end{equation}
where $n^{cb}$ and $n^{ub}$ are the number density of the ``charged baryons'' and  ``uncharged baryons'' respectively~\cite{Hu,Peng},
 and $n_s^{cb}$ and $n_s^{ub}$ are called scalar density for the charged and uncharged baryons~\cite{Ra2}. In $T = 0$\footnote{This can justified
 since the Fermi temperature of the neutron stars is too high compared to its own temperature~\cite{Sil}.}  they are given by:

\begin{equation}
dn^{ub} =  \frac{8\pi k^2}{(2\pi)^3} \quad \to \quad n^{ub} = \int_{0}^{k_f} \frac{8\pi k^2}{(2\pi)^3} = \frac{k_f^3}{3\pi^2}, \label{15}
\end{equation} 

\begin{equation}
dn^{cb} =\frac{|e|B}{(2\pi)^2} \eta{(\nu)}dk_z \nonumber ,
\end{equation}
\begin{equation}
n^{cb} =  \frac{|e|B}{(2\pi)^2}\sum_{\nu}^{\nu_{max}}\eta{(\nu)} \int_{-k_f}^{k_f}dk_z = \frac{|e|B}{2\pi^2}\sum_{\nu}^{\nu_{max}}\eta (\nu)k_f \label{16} ,
\end{equation}

\begin{equation}
n_s^{ub} = \frac{1}{\pi^2}\int_0^{k_f} \frac{M^{*}_b k^2 dk}{\sqrt{M^{*2}_b + k^2}}, \quad  \label{17}
\end{equation}
\begin{equation}
n_s^{cb} = \frac{|e|B}{2\pi^2} \sum_{\nu}^{\nu_{max}} \eta(\nu) \int_0^{k_f} \frac{M^{*}_b dk_z}{\sqrt{M^{*2}_b + k_z^2 + 2\nu|e|B}} . \label{18}
\end{equation}

The summation in $\nu$ in the above expressions ends at $\nu_{max}$, the largest value of $\nu$ for
which the square of Fermi momenta of the particle is still positive and which corresponds to the closest integer from below defined by the ratio:

\begin{equation}
\nu_{max} \; \le \; \frac{\mu^2 - M^{*2}_b}{2|e|B} , \quad \mbox{charged baryons}\label{19} 
\end{equation}

\begin{equation}
\nu_{max} \; \le \; \frac{\mu^2_l - m^{2}_l}{2|e|B} . \quad \mbox{leptons} \label{20}
\end{equation}

Now we couple the equations  imposing $\beta$  equilibrium and zero total net charge:

\begin{equation}
\mu_{b_i} = \mu_n -  e_i \mu_e, \quad  \mu_e = \mu_\mu,  \quad \sum_b e_b n^b + \sum_l e_l n^l =0 \label{21} ,
\end{equation}
where $\mu_{b_i}$ and $e_i$  are the chemical potential and electric charge of the i-th baryon, and $\mu_n$, $\mu_e$ and $\mu_\mu$
 are the chemical potential of the neutron, electron and muon respectively, $n^b$ is the number density of the baryons and $n^l$
is the number density of the leptons. 

The energy density of the neutron star is:

\begin{equation}
\epsilon  =  \sum_{ub}\epsilon_{ub} + \sum_{cb}\epsilon_{cb} + \sum_l \epsilon_l + \sum_m \epsilon_m + \frac{B^2}{8\pi} , \label{22}
\end{equation}
where the energy densities for the uncharged baryons, charged baryons, leptons and mesons have the following forms:

\begin{equation}
\epsilon_{ub} = \frac{1}{\pi^2}\int_{0}^{k_f} \sqrt{M^{*2}_b + k^2}k^2 dk, \label{23}
\end{equation}
\begin{equation}
\epsilon_{cb} = \frac{|e|B}{2\pi^2} \sum_{\nu}^{\nu_{max}}\eta(\nu) \int_{0}^{k_f} \sqrt{ M^{*2}_b +  k_z^2 +2\nu |e|B}dk_z \label{24} ,
\end{equation}
\begin{equation}
\epsilon_{l} = \frac{|e|B}{2\pi^2} \sum_{\nu}^{\nu_{max}}\eta(\nu) \int_{0}^{k_f} \sqrt{ m^{2}_l +  k_z^2 +2\nu |e|B}dk_z \label{25} ,
\end{equation}
\begin{equation}
\epsilon_{m} = \frac{1}{2}m_s^2\sigma_0^2 + \frac{1}{2}m_v^2\omega_0^2 + \frac{1}{2}m_\rho^2\rho_0^2 + \frac{1}{3!}\kappa \sigma_0^3 + \frac{1}{4!}\lambda \sigma_0^4 \label{26} .
\end{equation}

To find the pressure, we use the second law of thermodynamics, that gives a isotropic pressure:

\begin{equation}
p = \sum_i \mu_i n^{i} -  \epsilon + \frac{B^2}{8\pi} \label {27} ,
\end{equation}
where the sum runs over all fermions. Note that the contribution from electromagnetic fields  should be taken into account in the calculation of
 the energy density and the pressure.

\subsection{TOV equations and the density-dependent magnetic field}

The magnetic field of the surface of the magnetars are of order of $10^{15}G$, but can reach more than $10^{18}G$ in their cores.
 To reproduce this behaviour we use a density-dependent magnetic field given by~\cite{Pal,Ra,deb}:

\begin{equation}
B(n) = B^{suf} + B_0 \bigg [1 - exp \bigg \{-\beta { \bigg (\frac{n}{n_0} } \bigg )^{\alpha} \bigg  \} \bigg  ], \label{28}
\end{equation}
$B^{suf}$ is the magnetic field on the surface of the neutron stars, taken as
$10^{15}G$, $n$ is the total number density, $n= \sum n^b$,  $B_0$ is the
constant magnetic field. The two free parameters $\beta$ and $\alpha$ are
chosen to reproduce a weak magnetic field bellow the nuclear saturation
density, and a quickly growing one when $n$ $>$ $n_0$, in such a way that 
$B(n)>0.95B_0$ when $n=6n_0$. To reproduce this behaviour, we have set $\beta$ = $-6.5 \cdot 10^{-3}$ and $\alpha$ = 3.5.
 Now  $B$ is replaced by  $B(n)$ in the term $B^2/8\pi$ in our EoS.

To finish our analytical analysis, we write the TOV~\cite{TOV} equations:

\begin{equation}
\frac{dM}{dr} = {4\pi r^2 \epsilon (r)}, \label{29}
\end{equation}

\begin{equation}
\frac{dp}{dr} = -\frac{G\epsilon (r)M(r)}{r^2} \bigg [ 1 + \frac{p(r)}{\epsilon (r)} \bigg ] \bigg [1 + \frac{4 \pi p(r) r^3}{M(r)} \bigg ] \bigg [ 1 -\frac{2GM(r)}{r} \bigg ]^{-1} \label{30} ,
\end{equation}
which are the differential equations for the structure of a static, spherically symmetric, relativistic star in hydrostatic equilibrium. The equation of
 states developed in this work are used as input for these equations.  There are three minors problems with this approach. First, the pressure is not 
 really isotropic. Anisotropies arise due to the preferential direction  $z$ of the magnetic field.  Second, the energy of the magnetic field itself is a
 further source of gravitation, that may induce a gravitational
 collapse. Third, the gradient of the pressure is a source of
 repulsion,  which 
counter-balances gravity. So, if the magnetic field is strong enough,
the neutron star may blow up. We  discuss the validity of the
symmetric TOV equations more carefully  at the end of the paper.

\section{Results and discussion}

 We consider two families of neutron stars: one containing just protons, electrons and neutrons, which we call  ``Atomic Stars'' denoted by the letter $A$ in
 the legends, and other containing protons, electrons, neutrons, muons and hyperons, which we call ``Hyperonic Stars'' denoted by the letter $H$ in the legends.
 In the results we also include the crust of neutron star through the BPS EoS~\cite{pet}, but always taken into account the contribution of the magnetic field
 through the term $B(n)^2/8\pi$ in the EoS. 

\noindent We choose two values for the magnetic field:  1.0 $\cdot$ $10^{17}{G}$ and 3.1 $\cdot$ $10^{18}{G}$  to produce a weak and a strong influence.
 We also include here some theoretical and observational constrains. First, all our EoS are causal and obey the Le Chantelier
 principle, \textit{i.e.}, the quantity $dp/d\epsilon$ lies between 0 and 1.   We plot the numerical results of  four   EoS  in  fig. 1.
%
\begin{figure}[ht]
\centering
\includegraphics[angle=270,
width=0.8\textwidth]{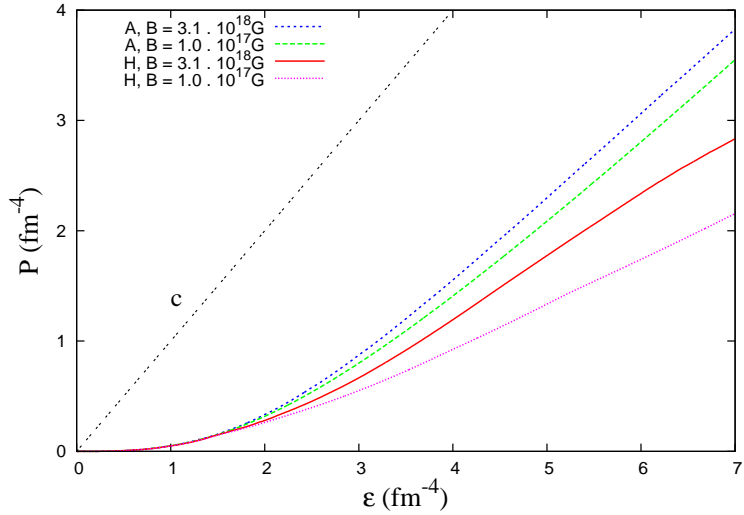}
\caption{EoS for two atomic and two hyperonic stars obtained with different values of the magnetic field.
The  straight line corresponds to the causal limit, for which $\epsilon = p$.}
\end{figure}

As we can see from   fig. 1, the presence of hyperons softens the EoS more than the influence of the magnetic field can stiffen it. No matter how strong is
 the magnetic field in the interior of the magnetar, the EoS of an atomic star
 is always stiffener than a hyperonic one. We can also see that all our EoS are causal
 (the ``c'' line is the causality limit).

From  figs. 1 and  3 we note that both the EoS and the fraction of particles $Y_i=n_i/n$ are not affected significantly by  a magnetic field about $10^{17}G$.
 The reason is that a field of this magnitude is too weak to contribute to the final EoS and to the fraction of particles. We also see that the magnetic field
 affects more the hyperonic stars than the atomic ones. As the hyperonic stars are  softer than the atomic ones, they are therefore  more sensitive to the presence
 of the magnetic field. Also the hyperonic stars are  denser than the atomic stars, with a bigger central density $n_c$. Moreover, for a fixed value of density,
  the pressure of the hyperonic stars are smaller than the atomic ones. So, as the magnetic field couples to the number density through eq. (\ref{28}),
 the contribution of the magnetic field  is always  greater in hyperonic than in atomic stars. We show this result plotting the pressure $p$ in function of $n$ in
  fig. $2$. 

\begin{figure}[ht]
\centering
\includegraphics[angle=270,
width=0.8\textwidth]{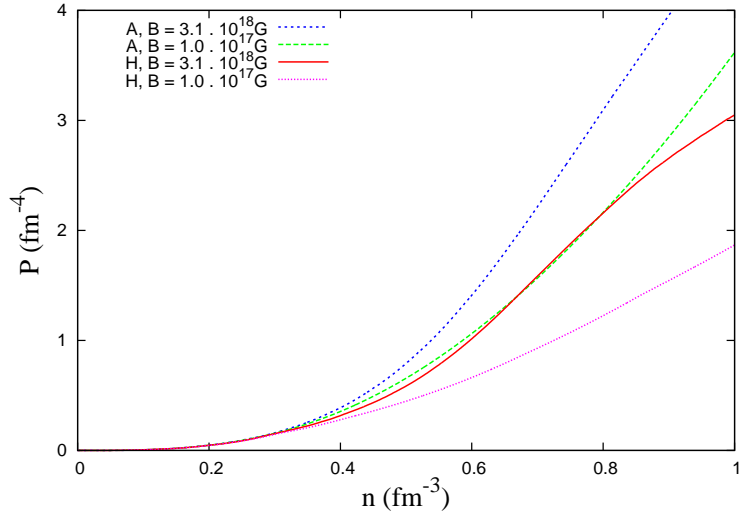}
\caption{Pressure as function of number density for the four stars. The relative difference between blue-green (red-rose) lines for atomic (hyperonic) stars  is the 
direct influence of the magnetic field.}
\end{figure}

 Figures 3 and 4 show the fraction of particles. We can see that for
 the strong field, the appearance of charged particles is favoured at low densities due
 to their dependence on the magnetic field, as expected from
 eq. (\ref{16}).   The behaviour of the particles in the presence of
 a strong magnetic field is also altered. For weak fields, the
 population of a kind  of particle is always well-behaved, while for the strong one many kinks appear.
The reason is that in the absence of a magnetic field the number density of a determined kind of particle grows smoothly with the momentum.
 When a magnetic field is present, there is also  a dependence of the discrete $LL$. For  a weak magnetic field a lot of Landau levels are occupied,
 but for a strong magnetic field, just a few of them are filled. So the orbit normal to the $z$ direction is tightly quantized. This effect is
more evident in the hyperonic stars.
 Each nozzle in a slope of a determined particle indicates that the density is high enough to create another Landau level. For a high density
 there are so many Landau levels available that the distribution approaches to the continuous, while for a weak magnetic field, even in a low density
 there are several $LL$, so there is no significant difference with the case in the absence of the magnetic field. One can notice that for
 hyperonic stars, at densities of the order of $0.8fm^{-3}$ the neutron is no longer the most important constituent. From this point, the $\Lambda^{0}$
 hyperon dominates in the region of high densities.

 We can also ask how the magnetic field affects the strangeness
 fraction, defined as
\begin{equation}
 f_s= \frac{1}{3} \frac{\sum_j |s_j|n_j}{n} , \label{30b}
\end{equation}
where $s_j$ is  the strangeness of baryon $j$ and $n$ is the total
number density. We  plot the results in  Fig. 5.

\newpage
 
 \begin{figure}[ht]
 \centering{Fraction of particles $Y_i$ as a function of number density $n$.}
\begin{tabular}{cc}
\includegraphics[width=5.9cm,height=7.cm,angle=270]{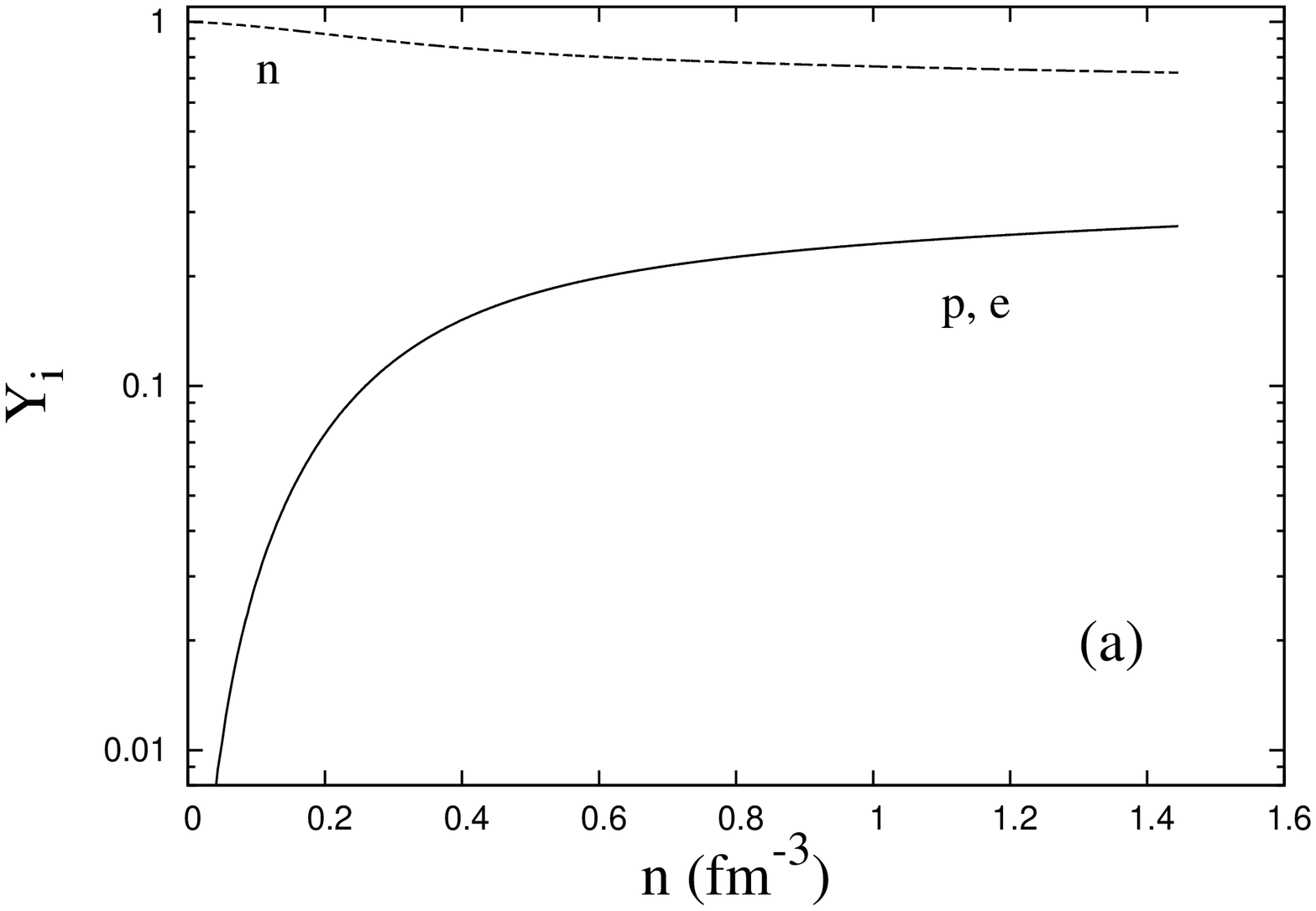} &
\includegraphics[width=5.9cm,height=7.cm,angle=270]{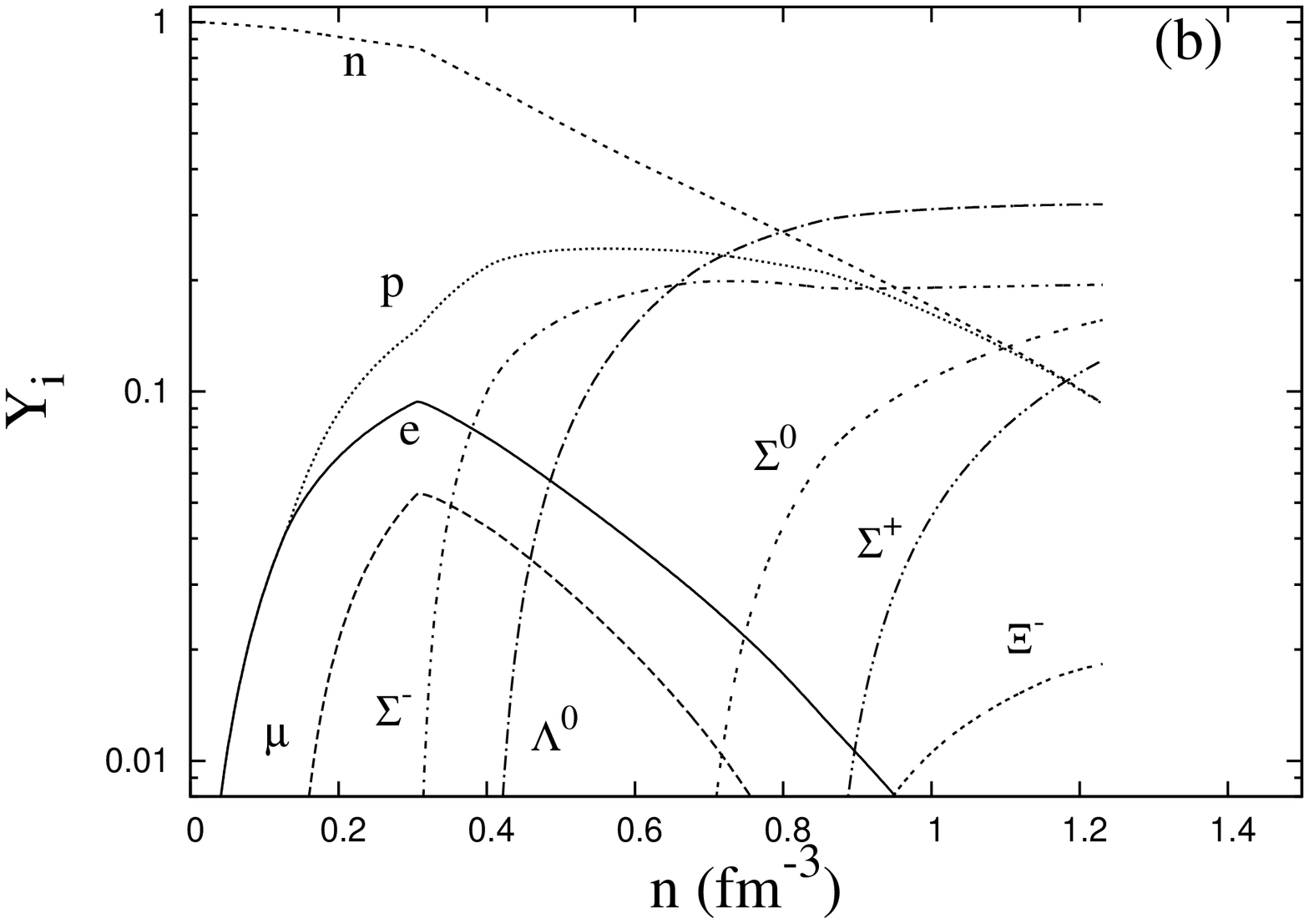} \\
\end{tabular}
\caption{ $(a)$ atomic stars and  $(b)$ hyperonic stars for a magnetic field of 1.0 . $10^{17}G$}
\end{figure}
\begin{figure}[ht]
 \centering
\begin{tabular}{cc}
\includegraphics[width=5.7cm,height=7.cm,angle=270]{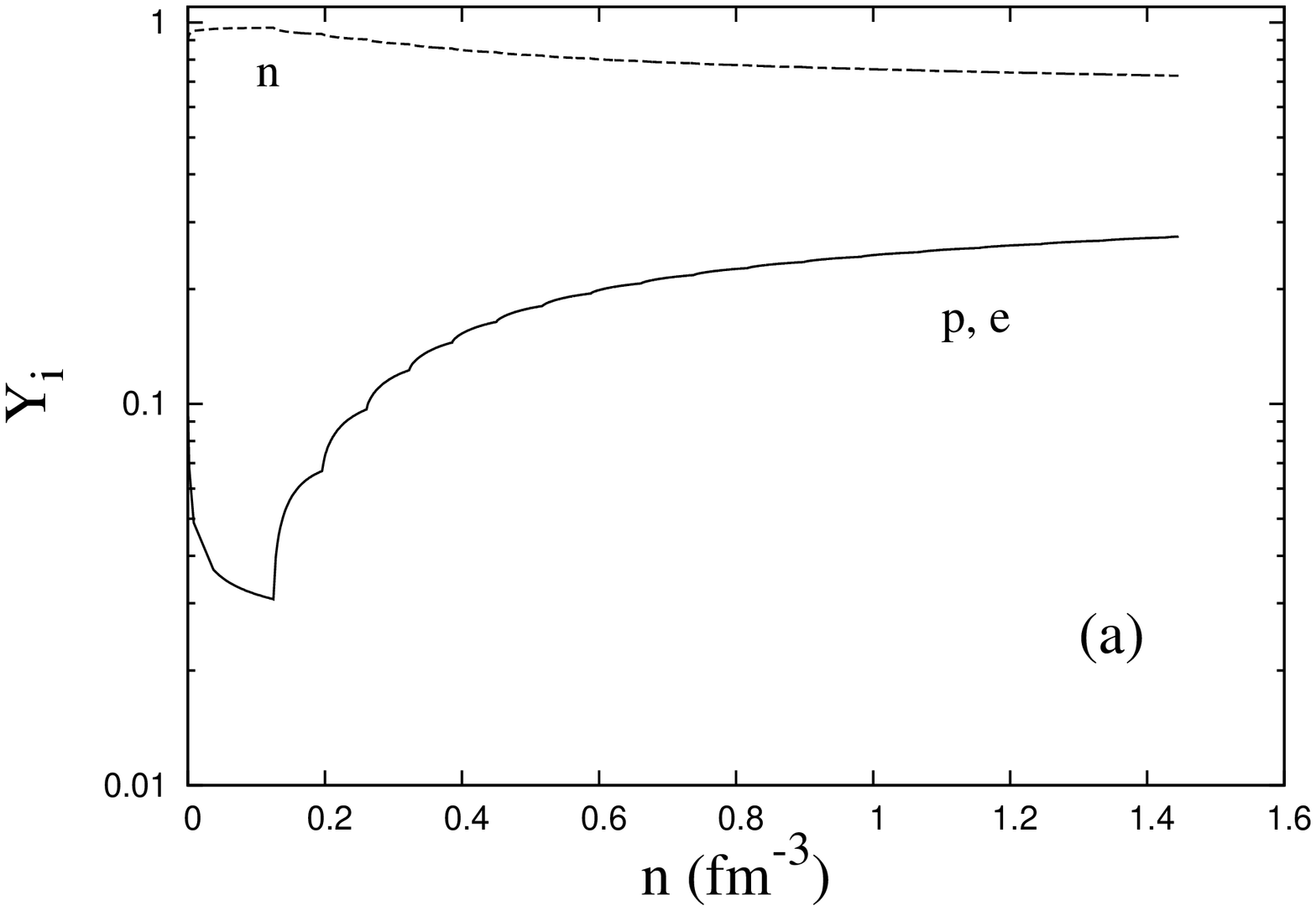} &
\includegraphics[width=5.7cm,height=7.cm,angle=270]{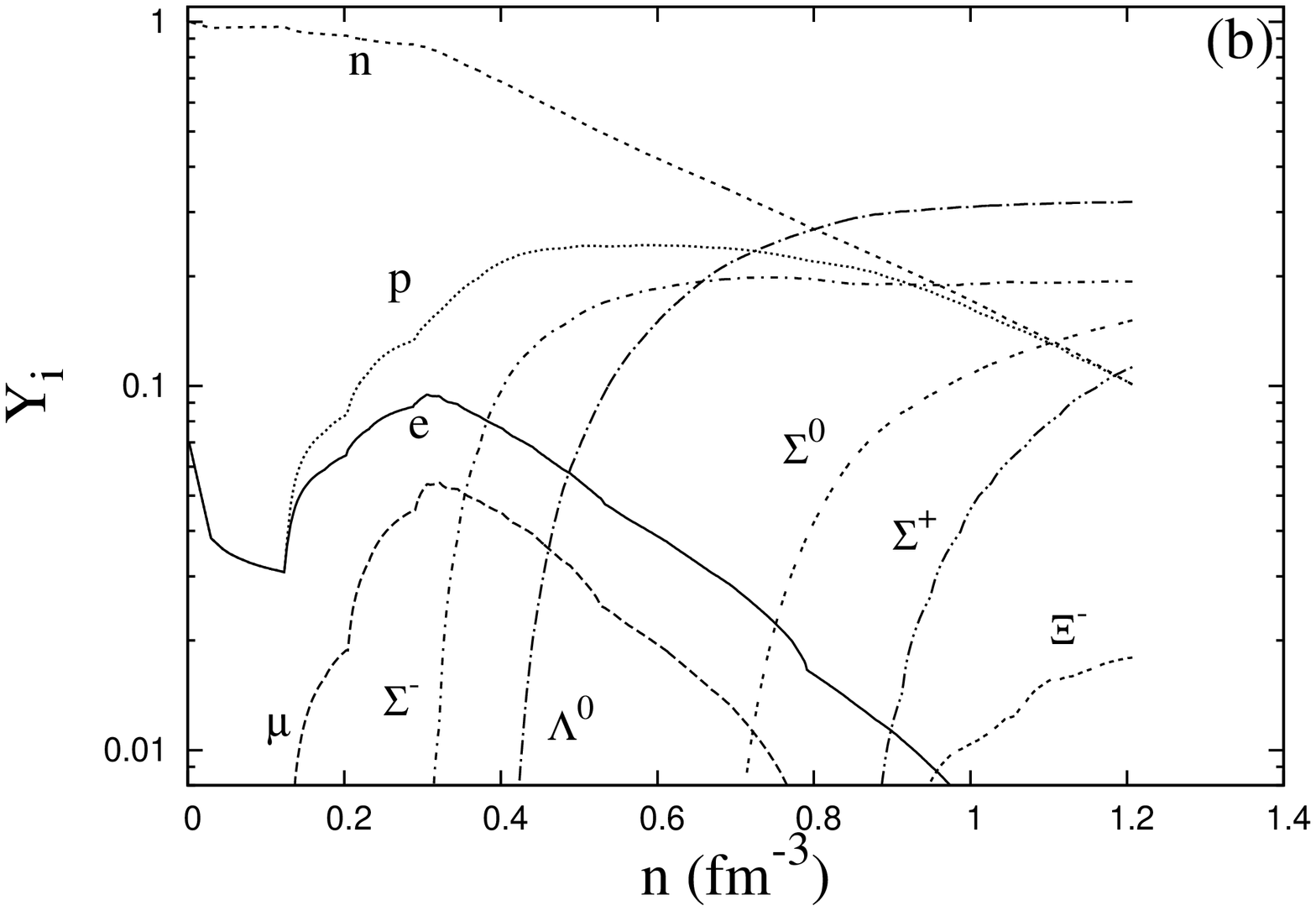} \\
\end{tabular}
\caption{ $(a)$ atomic stars and  $(b)$ hyperonic stars for a magnetic field of 3.1 . $10^{18}G$}
\end{figure}

\begin{figure}[ht]
\centering
\includegraphics[angle=270,
width=0.8\textwidth]{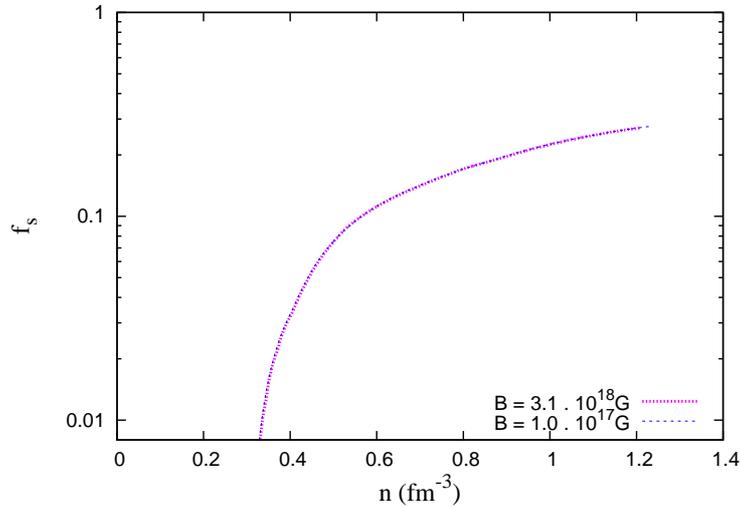}
\caption{The strangeness fraction for a weak and a strong magnetic field.}
\end{figure}

As we can see, although the magnetic field causes a difference in the
fraction of a single particle,  no difference is found in the total strangeness fraction.
For a fixed density, if a new particle is created, another has to
disappear, due to the $\beta$ equilibrium condition. Hence, although the Landau quantization 
affects a single particle, this effect is washed out when all of them
are summed.
 
 It is hard to compare quantitatively our results with other ones existing in the literature due to  the many different  parametrizations.
 In the absence of hyperons, we see that our EoS is stiffener than
 those presented in~\cite{Ra2,Bro} since the authors of those 
references do not consider
 a density-dependent magnetic field. With relation to the fraction of particles $Y_i$, the fraction of protons in a low density region is more
 favoured in~\cite{Ra2} than in our work. Now for a magnetic field of 1.0 $\cdot$ $10^{17}G$ there is virtually no difference compared with the results
 with zero magnetic field presented in~\cite{deb2}\footnote{ The same holds for the TOV solution showed bellow.}.

 When a strong magnetic field is applied, we see that our
 parametrization do not prevent the hyperon formation in a 
low density region as showed in~\cite{Aziz}.
 Also, in our work, the influence of Landau quantization is much  more evident even for a smaller value of the magnetic field.

In order to validate our EoS, we have to solve the TOV equations and
check if the results agree with observational constraints. But, first we have to check
 if the TOV equations are allowed  to be used as a good
 approximation. The first point is that the pressure is not really isotropic. However, recent work~\cite{XH}
 showed that the anisotropy is not significant till the magnetic field
 reaches $B \simeq 3.2 \cdot 10^{18} G$. This is the reason  we use
 the value  $3.1 \cdot 10^{18}G$ as the strongest magnetic field.  The second concerns the magnetic field as a source of
 gravitational energy. We expect that the magnetic field induces gravitational collapse.
 A simple way to avoid it, is by requiring that the energy of the
 strongest magnetic field, $B_0 = 3.1 \cdot 10^{18}G$, do not be dominant. In other words,
 we  require that  $ \epsilon_M > B(n)^2/8\pi$, where $\epsilon_M$ is the energy density of  matter, given by  eq. ($\ref{22}$) without the last term,
 which is the energy of the magnetic field itself.  We can define the dimensionless  quantity $co$:

\begin{equation}
 co =  \bigg ( 1 -  \frac{B(n)^2}{8\pi \epsilon_M} \bigg ) \label{31}, 
\end{equation}
and require that $co$\footnote{We have named $co$ and $bo$ after their
  relation with the collapse and the bound state respectively.} is  never negative.
 As we can see from the fig. 6 this imposition is fully filled.

\begin{figure}[ht]
\centering
\includegraphics[angle=270,
width=0.8\textwidth]{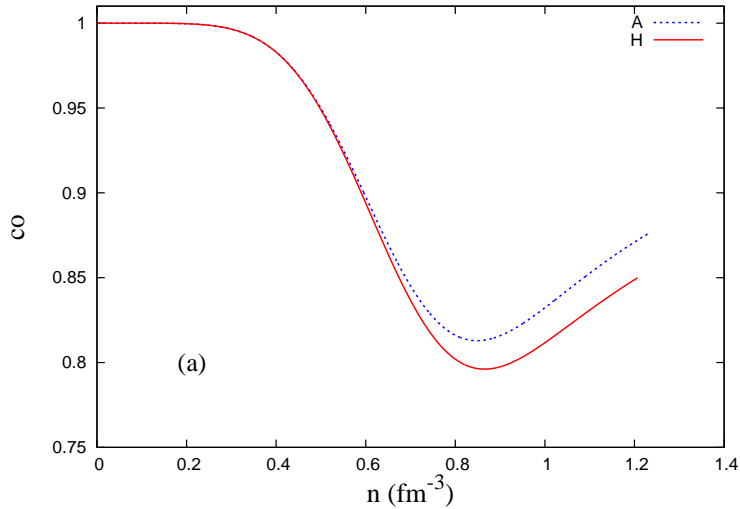}
\caption{The co quantity, showing that the energy density of the magnetic field is not dominant.}
\end{figure}

The last point refers to the pressure of the magnetic field. If the magnetic field is too strong,  the neutron star cannot be gravitationally bounded.
In order  to study the limits of the magnetic field such that it does not blow up the star we use the approximation of radial magnetic field  presented
in~\cite{MANE}. For the neutron star to be gravitationally bound, the term $M(r) + 4\pi r^3(p - B(n)^2/8\pi)$ - where $p$ here is the pressure of matter  
given by the eq. ($\ref{27}$) without the term of the magnetic field
itself - has always to be positive\footnote{We do not use 
this modified TOV equations since they imply the existence of a
magnetic monopole.}. 
So, if the term $p - B(n)^2/8\pi$ is positive,
of course the total sum also is. Now we define another dimensionless quantity $bo$:

\begin{equation}
 bo =  \bigg ( 1 -  \frac{B(n)^2}{8\pi p} \bigg ) \label{32}, 
\end{equation}
and require that $bo$ is always positive. From  Fig. 7 we see that this requirement is reached, so we guarantee that the neutron star is gravitationally bound
for all fields used in this work.

\begin{figure}[ht]
\centering
\includegraphics[angle=270,
width=0.8\textwidth]{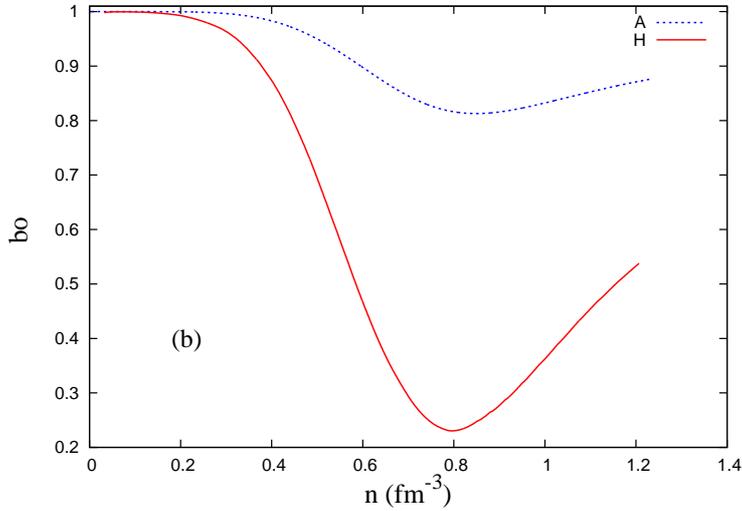}
\caption{The bo quantity, showing that the neutron star is always gravitationally bound.}
\end{figure}

 Now we can return to the TOV constraints.The star masses cannot exceed the maximum theoretical neutron star mass of 3.2 solar masses~\cite{ruf}.
 The EoS has to be able to predict the 1.97 solar  masses neutron star~\cite{demo} and to be in agreement with the redshift measurements ($z$) of
 two neutron stars. A redshift of $z=0.35$ has been obtained  from three different transitions of the spectra of the X-ray binary EXO0748-676~\cite{Cottam}.
 This redshift corresponds to M/R=0.15 $M_{\bigodot}$/$km$.  Another constraint on the mass-radius ratio comes from the observation of two
 absorption features in the source spectrum of the 1E 1207.4-5209 neutron star,  with redshift from $z=0.12$ to $z=0.23$,  which gives M/R=0.069 $M_{\bigodot}$/$km$
  to M/R=0.115 $M_{\bigodot}$/$km$~\cite{Sanwal}. Besides these constrains,  the stars with central density above that of the maximum mass stars
 are mechanically unstable~\cite{Glen}. Due this fact, the $\Xi^{0}$ hyperon is not present  in the neutron star interior
 (the density required to create it is too high\footnote{Indeed the $\Xi^{0}$ hyperon appears in an insignificant quantity  $Y_{\Xi^{0}}$ = $10^{-5}$ at 1.2$fm^{-3}$.}). 

Solving the TOV equations for the  EoS   we obtain the results
presented in Table 2.

\begin{table}[ht]
\centering
\scalefont{.92}

\caption{Neutron stars properties computed from the four
EoS used as input to the TOV equation. From the central density $n_c$, 
we see that the hyperonic stars are denser than the atomic ones for a fixed magnetic field.
 $B_0$ and $B_c$ are given in multiple of $10^{17}G$ }
\label{tab:2}       
\begin{tabular}{lllllll}\hline\noalign{\smallskip}
  Kind &  M/$M_{\odot}$ (Max.) & R (km) & $n_c$ ($fm^{-3}$) & $B_0$ & $B_c$   \\
\noalign{\smallskip}\hline\noalign{\smallskip}
 A & 2.48 & 12.21 & 0.741  & 31.0 & 25\\
\noalign{\smallskip}\hline\noalign{\smallskip}
A  & 2.39 & 12.10 & 0.840  & 1.0 & 1.0\\
\noalign{\smallskip}\hline\noalign{\smallskip}
 H  & 2.22 & 11.80 & 0.824  & 31.0 & 28 \\
\noalign{\smallskip}\hline\noalign{\smallskip}
H  & 2.01 & 11.86 & 0.952  & 1.0 & 1.0 \\
\noalign{\smallskip}\hline
\end{tabular}
\vspace*{1cm}  
\end{table}

The fact that the EoS of hyperonic stars are always  softer than the atomic ones reflects in the maximum mass of these stars.
 Further,  from Table 2 we can see that the hyperonic stars are denser
 than the atomic ones for a fixed magnetic field, as stated before. 
Moreover, denser stars can support stronger magnetic fields in their center ($B_c$) as
a consequence of eq. (\ref{28}).  A curious fact is that  while the radius of the
hyperonic star with the maximum mass decreases for strong magnetic
fields, in the atomic stars the radius grows with the increase of
$B$. This result is a consequence of the fact that
the radii of the neutron stars are not related to the stiffness of the EoS but with its symmetry energy slope~\cite{Rafa}.
 We plot the TOV solutions in Fig. 8.

\begin{figure}[ht]
\centering
\includegraphics[angle=270,
width=0.8\textwidth]{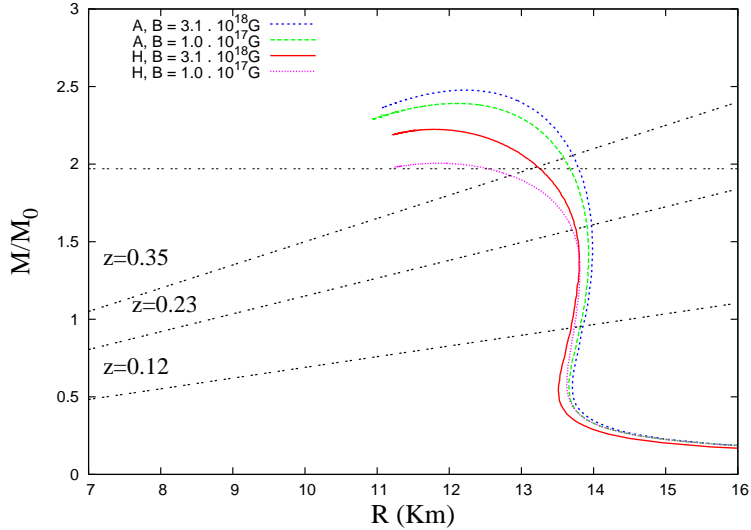}
\caption{Mass-radius relation for two atomic and two hyperonic stars with different values of magnetic
field. The straight lines are the observational constraints.}
\end{figure}

We can see that all our models are in agreement with the previously discussed constrains~\cite{ruf,demo,Cottam,Sanwal}. The inclined straight
 lines are the constrains of the measured redshift while the horizontal one is the 1.97 $M_{\bigodot}$ pulsar. Every single dot in the curves is
 a possible neutron star. We can see also that a magnetar can reach a mass of
 2.48$M_{\bigodot}$. However, as we neither know, nor expect to discover
 any white dwarf with mass above 1.4$M_{\bigodot}$ due to the Chandrasekhar 
limit~\cite{shap}, the same holds to the neutron stars
 with masses above 2.0$M_{\bigodot}$ due to the Oppenheimer-Volkoff limit if 
we believe that hyperons and muons exist in their interior.
 The maximum possible mass is very close to the 1.97 $M_{\bigodot}$  known 
neutron star. The discovery of neutron stars with masses above
 2.0$M_{\bigodot}$ will imply that either hyperons are absent or the star is a  
rare strongly magnetized neutron star, as shown in fig. 8.

We have seen from Fig. 5 that the strangeness fraction $f_s$ is not
affected by the magnetic field for a fixed density.
Let's see what happens inside different stars with different central
densities, which can support different magnetic fields.
We first calculate the
strangeness fraction in the inner core by fixing the mass, and check what happens with
 $f_s$ when we vary the magnetic field. We then fix the maximum mass
 and look at its strangeness content.  We write these results in the Table 3.
We see that the strangeness fraction is not significant for lower mass
stars. Indeed, the hyperons just become important for stars with
masses above 1.6 $M_\odot$. 
The magnetic field is also not important for stars with low masses due
to the low magnetic field strengths reached in their center. 
For masses above $1.6M_{\odot}$  both the magnetic field and  the 
hyperonic constituents become important.
 Since the magnetic field reduces the central density, it suppresses
 the hyperon formation. 
To conclude  this discussion we note that the$f_s$ is the value calculated
 at the center of the neutron stars, so the quantity of hyperons  at
 lower densities is even more  insignificant.

\begin{table}[ht]
\centering
\scalefont{.98}

\caption{The influence of the magnetic field in the strangeness fraction for some standard Mass.
The B's are given in multiple of $10^{17}G$}
\label{tab:3}       
\begin{tabular}{llllll}\hline\noalign{\smallskip}
 M ($M_{\bigodot}$) & $f_{s}$ & $n_c$ ($fm^{-3}$) & $B_0$ & $B_c$  \\
\noalign{\smallskip}\hline\noalign{\smallskip}
1.2 & 0.00 & 0.29  & 31.0 & 1.8 \\
\noalign{\smallskip}\hline\noalign{\smallskip}
1.2 & 0.00 & 0.29  & 1.0 & 0.07 \\
\noalign{\smallskip}\hline\noalign{\smallskip}
1.4 & 0.01 & 0.32  & 31.0 & 2.6 \\
\noalign{\smallskip}\hline\noalign{\smallskip}
1.4 & 0.01 & 0.34  & 1.0 & 0.12 \\
\noalign{\smallskip}\hline\noalign{\smallskip}
1.6 & 0.03 & 0.38  & 31.0 & 4.5 \\
\noalign{\smallskip}\hline\noalign{\smallskip}
1.6 & 0.04 & 0.40  & 1.0 & 0.22 \\
\noalign{\smallskip}\hline\noalign{\smallskip}
1.8 & 0.05 & 0.44  & 31.0  & 7.2\\
\noalign{\smallskip}\hline\noalign{\smallskip}
1.8 & 0.08 & 0.52  & 1.0 & 0.40 \\
\noalign{\smallskip}\hline\noalign{\smallskip}
\noalign{\smallskip}\hline\noalign{\smallskip}
2.22 (Max) & 0.18 & 0.824  & 31.0 & 28 \\
2.01 (Max) & 0.23 & 0.952  & 1.0 & 1.0 \\
\noalign{\smallskip}\hline
\end{tabular}
\vspace*{1cm}  
\end{table}

To conclude this paper we  discuss  the cooling of the neutron stars
due the direct URCA effect. As pointed out previously, Ref.~\cite{LPPH} showed that
the neutron stars can to cool much faster, if their proton fractions
reach values of $\simeq 11- 15\%$. If it occurs, the direct URCA process enhances neutrino
 emission, cooling the neutron star in a rate much faster than any
 other process.  We next show the strangeness fraction of the hyperonic stars
 when the direct URCA process  is allowed, by writing the mass of the
 stars when the proton fraction reaches 13$\%$
 for atomic (A) and hyperonic (H) stars in table 4. 

\begin{table}[ht]
\centering
\scalefont{.95}

\caption{Lower mass of the neutron star for a proton fraction of $13\%$.}
\label{tab:4}       
\begin{tabular}{lllllll}\hline\noalign{\smallskip}
 M ($M_{\bigodot}$) & Kind & $f_s$ & $n_c \; (fm^{-3})$ & $B_0$ & $B_c$  \\
\noalign{\smallskip}\hline\noalign{\smallskip}
1.50 & A & - &   0.33  &  31.0 & 2.9 \\
\noalign{\smallskip}\hline\noalign{\smallskip}
1.46 & A & - & 0.33 & 1.0 & 0.11 \\
\noalign{\smallskip}\hline\noalign{\smallskip}
1.12 & H & 0.00 & 0.27 & 31.0 & 1.4 \\
\noalign{\smallskip}\hline\noalign{\smallskip}
1.08 & H & 0.00 & 0.27 & 1.0 & 0.06 \\
\noalign{\smallskip}\hline\noalign{\smallskip}
\noalign{\smallskip}\hline
\end{tabular}
\vspace*{1cm}  
\end{table}

We see that the magnetic field causes almost no change in the lower mass that enables direct URCA process.
On the other hand, the presence of muons (since hyperons are not
present in low densities) has a great influence on the lower mass 
when the proton fraction reaches $13\%$.

\section{Conclusion}

In this work we consider a hadronic neutron star composed by hyperons submitted to a strong magnetic field.
 We  see that while the presence of hyperons reduce the maximum mass by softening the equation the state
 (EoS)~\cite{Glen,deb2,Glen2,Ma,Z,HD}, the presence of a density-dependent magnetic field tends to increase the
 maximum mass stiffening the EoS~\cite{Ra,deb,Aziz}. Our  study shows also that although the influence of a
 strong magnetic field stiffener the EoS, a hyperonic magnetar still has a softer EoS compared with a common
 atomic stars.  We  also see how the magnetic field can change the chemical composition of neutron stars due
 the Landau quantization and  offer an explanation about the non-existence of  
neutron stars  with masses above the  1.97 $M_{\bigodot}$. We have
shown that the TOV equations can be used 
as an approximation for neutron stars subject to
strong magnetic fields up to a certain limit
 and that the effect of strong magnetic fields has to  be taken into account in a description of massive magnetars,
 mainly if we believe that there are hyperons in their interior.  In
 this case, the magnetic field increases the mass in more than $10\%$ although
its effect is not significant for low mass neutron stars because the central magnetic field $B_c$ is also low.
 According to the present model and parametrization, it is unlikely that hyperons
 are present in neutron stars with masses below $1.4M_\odot$ even if
 the magnetic field is considered.

\vspace{.5cm}

$\mathbf{Acknowledgements}$

\vspace{.5cm}

 This work was partially supported by CNPq and CAPES.

%

%
%

\end{document}